\documentstyle[epsfig,twocolumn]{mn}
\def\div{\mathop{\rm div}}
\def\curl{\mathop{\rm curl}}

\def\be{\begin{eqnarray}}
\def\ee{\end{eqnarray}}

\newcommand {\B}{\bmath{B}}

\newcommand {\vv}{\bmath{v}}
\title[Vortex-interface interactions]{Vortex-interface interactions 
and generation of glitches in pulsars}
\author[A. Sedrakian and J. M. Cordes]
{Armen Sedrakian and  James M. Cordes\\
Center for Radiophysics and Space Research, 
Cornell University, Ithaca, NY 14853, USA}
\date{Accepted 1999 February 11. 
     Received 1999 January 7; in original form 1997 December 8
     }
\pagerange{1-12}
\volume{000}
\pubyear{1999}

\def\LaTeX{L\kern-.36em\raise.3ex\hbox{a}\kern-.15em
    T\kern-.1667em\lower.7ex\hbox{E}\kern-.125emX}

\begin{document}

\maketitle
\begin{abstract}
\noindent
We show that the crust-core interface in neutron stars acts as a potential 
barrier to the peripheral neutron vortices approaching the interface
in the model in which these are coupled to the proton vortex clusters.
This elementary barrier because of the interaction of vortex 
magnetic flux with the  Meissner currents set up by the crustal 
magnetic field at the interface. The dominant part of the 
force is derived from  to the cluster-interface interaction. 
As a result of the stopping of the continuous neutron vortex current through
the interface, angular momentum is stored  in  the superfluid layers 
in the vicinity of the crust-core interface during the interglitch period.
Discontinuous annihilation of proton vortices on the boundary restores 
the neutron vortex current and spins up the observable crust on short 
time-scales, leading to a  glitch in the spin characteristics of a pulsar.
\end{abstract}


\section{Introduction}

\subsection{Motivation}

The many-body calculations of the ground state energy of the matter
in  neutron stars imply an internal structure 
represented by succession of phase transitions as one moves 
from the surface to higher densities. The outer envelope of
highly compressed solid melts into a homogeneous
neutron-proton-electron Fermi liquid at roughly half the nuclear 
matter saturation density. The phase transition is 
an analog of the liquid-solid phase transition, 
is of the first order, and is signalled by an instability of the continuum
proton liquid in the core against clustering into heavy nuclei in 
the crust (Pethick, Ravenhall \& Lorenz 1995). 
In our subsequent discussion of the vortex-interface interaction we 
shall mainly focus on the crust-core interface and shall identify 
its location according to this criterion. 
The internal interface bounding the superfluid, superconducting neutron-proton
liquid at supranuclear densities may correspond to either disappearance 
of the superfluidity of the nucleonic matter or onset of  
a kaon-condensed phase (e.g. Brown, et al 1994; and Pandharipande, Pethick 
\& Thorsson 1995) or deconfined quark plasma (e.g. Glendenning 1996 and 
references therein). In view of the uncertainties involved in the physics of 
high-density nuclear matter, we shall only briefly discuss the vortex 
interactions with internal interfaces.

The density profiles of the  superfluid phases in a neutron stars 
do not coincide with the profiles of various phases of
dense  matter in general. The boundaries of the latter phases thus 
may play an important role in the dynamics of superfluids 
which support  the vortex lattice state. 
For example, the rotation of neutrons is supported by a Feynman-Onsager 
vortex lattice state and their interaction with the phase boundaries 
can affect their dynamics; the same applies to the proton vortex lattices,
carrying the magnetic flux through the superconducting proton liquid.

In the case of laboratory superfluids the  
vortex-interface  interactions have been 
studied  both experimentally and theoretically. 
An example is the onset of the resistive state in
type-II superconductors when a current passes perpendicular to the 
applied magnetic field. The value of the critical current of sufficiently clean
samples, where the pinning effects are negligible, remains finite, and 
is determined by the  Bean-Livingstone barrier acting on 
the vortex lattice at the boundary. Experimental measurements
on Nb by Lowell (1968) confirmed the existence of the barrier;
a theoretical discussion can be found, for example, in de Gennes (1966).
An example of manifestation of vortex-interface interaction  
in the neutral superfluids is the dependence of the lower critical 
velocity of vortex  nucleation in superfluid $^4$He on
the roughness of the vessel inner surface (see e.g. Sonin \& Krusius 1994).

In this paper we derive the vortex interface intercations occurring at the 
boundaries of the superfluid phases in neutron stars,
and examine  the conditions under which 
this type of interaction could be responsible for the phenomenon of
glitches in the spins of pulsars. Any theoretical model should explain 
the following observational facts:
(i) short spin-up time-scales, which are observed to be less 
than 120~s in the Vela pulsar B0833-45 and less than an hour in the 
Crab pulsar B0531+21; (ii)  magnitudes of the jumps in the rotation 
and spin-down rates, $\Delta\nu/\nu \sim 10^{-8}-10^{-6}$ and 
$\Delta\dot\nu/\dot\nu \sim 10^{-3}-10^{-2}$, respectively; 
and (iii) the origin of the instability  driving a glitch, along 
with characteristic intervals between glitches for a given pulsar; 
the latter  times-cales range from several months to several years 
depending on the object.
The items (i)-(iii) will be addressed here  
using the vortex cluster model for the superfluid 
core dynamics (Sedrakian et al 1995a,b hereafter Papers I and II).

The physical picture is the following. 
In the interjump epoch a neutron star is  
decelerating; consequently the vortex lattice in the superfluid core is 
expanding and the peripheral vortices attempt to cross the crust-core 
boundary. The crust-superfluid core interface acts as a 
potential barrier to the proton vortices in the superfluid core 
that approach this boundary. If the repulsive component 
of the force, derived from the potential,  is added to the 
force balance condition for the neutron vortices, it gives 
rise to an imbalance along  the radial direction - 
the component of the friction force 
in this direction drops to zero, while the effective Magnus force becomes
proportional to the difference between the velocities of the normal and 
superfluid components. The deceleration of the star will lead  to a growth
of this force until  the proton vortices are able to annihilate 
at the interface. The neutron vortices will 
relax to their equilibrium positions,
imparting angular momentum 
from the superfluid to the normal component thus, producing 
a pulsar macrojump. This picture is akin to the unpinning 
model of Anderson \& Itoh (1975),
except that the interaction does not involve the bulk of the 
superfluid, but rather its layer at the phase boundary. 
The latter setup has the advantage of a coherent onset of 
the glitch and does not require  repinning.
A brief sketch of this macrojump generation mechanism has been given 
elsewhere (Sedrakian \& Cordes 1998).

\subsection{Overview}

The first suggested mechanisms for triggering the glitches were based on the 
idea of starquakes (Ruderman 1969; 
Smoluchowski \& Welch  1970; Baym \& Pines 1971; Carter \& Quintana 1975).
These models  employed the idea of discontinuous readjustments of 
the  shape of the  star as it spins down under external dissipative 
torques. However, the possibilities of the 
explanation of the recurrence rate of glitches in the Vela and 
Crab pulsars within these models are severely limited; 
e.g. Baym \& Pines (1971) find that the recurrence time
is   $\sim 10^8$ yr and $\sim 10^4$ yr for the 
Vela and Crab pulsars, respectively, if these are assumed to be 
1.4 $M_{\odot}$ neutron stars.

Instabilities associated with the superfluid 
component were proposed as being 
caused by long-living persistent currents and sudden 
annihilation of excess vortices (Packard 1972). A specific mechanism 
for the occurrence of the instability was proposed by Anderson \& 
Itoh (1975), who argued that glitches occur through the 
pinning and unpinning of vortices in neutron star 
crusts [see also Anderson et al (1982) and Itoh (1983)]. 
The required moment of inertia that can be accumulated in the 
interglitch period and the mechanisms for the unpinning are    
suitable for the large glitches observed 
in Vela-type pulsars (Alpar  et al 1981, 1993, Link \& Epstein 1991). 
However the subsequent repinning, 
which is required for the recurrence of glitches 
in this model, is still not well understood  (Shaham 1980, 
Sedrakian 1995). Jones (1991a,1993) argued that the superfluid
is not pinned in the whole bulk of the crusts, and the discontinuous
relocation of the interface between the 
phases with pinned and unpinned vortices 
can trigger a glitch. Ruderman (1991) suggested that a
spinning-down superfluid neutron star would 
strain the crust beyond its elastic yield strength, which would lead 
to crust cracking, resulting in glitches with magnitude and recurrence rate 
compatible with those observed. Thermal effects as a cause of a glitch 
were  discussed by Greenstein (1979). 
Sudden perturbation of the inner crust 
temperature, leading to an increase of the frictional coupling between the 
superfluid and the crust,  have beed simulated by Link \& Epstein (1996), 
who find  spin-ups compatible with the glitches in the Crab and Vela pulsars. 
The effects of the  exotic nuclear structure environment  of the inner crust 
(e.g.  Lorenz, Revenhall \& Pethick 1993) on the vortex 
dynamics have been considered 
by  Mochizuki,  Oyamatsu \&  Izuyama (1997), who find that 
nuclear rod structure can pin a vortex line and can
be an origin of vortex accumulation, leading to a glitch in the spirit
of the Anderson-Itoh model.

Note that the glitch generation mechanisms, apart from describing 
the transients, lend themselves as candidates for generation  
of the sustained spin fluctuation processes manifested as `timing noise' 
(Cordes \& Greenstein 1981, Cordes, Downs \& Krause-Polstorff 
1988 and references therein). Simple amplitude down-scaling of the transients
is not sufficient, however, because the transients are biased towards one 
sign of the spin-fluctuation (e.g. spin-up), while the timing noise
fluctuations require deviations of both signs.   

\subsection{Organization of the paper}

In Section 2 we derive general expressions for the vortex-interface
interaction in  three-velocity superfluid hydrodynamics and give 
order-of-magnitude estimates of the interaction strength 
at the crust-core interface. In Section 3 the collective interaction 
effects - the cluster-interface interaction - are discussed and a comparison 
with the Magnus force needed to build-up an instability is made.
The microscopic time-scales relevant for the 
spin-up problem are considered in Section 4. Section 
5 is a summary of our results.

\section{Vortex-Interface Interactions in Three-Velocity Hydrodynamics}

We shall focus further on the dynamics of the superfluid outer core, 
where both the neutrons and the protons are in the superfluid state, 
and apply the Newtonian version of the superfluid hydrodynamics 
[a detailed exposition of the theory  and references to earlier work 
can be found in Mendell \& Lindblom (1991), Mendell (1991), and Paper I]. 
The relativistic counterparts of superfluid hydrodynamic equations 
have been made available recently (Langlouis, Sedrakian, Carter 1998,  
Carter \& Langlouis 1998); their application to the problem 
at hand, however, is not straightforward and we shall continue to 
work entierly in the Newtonian limit. 

To keep the discussion general, the ground state structure of 
vortex lattices will not be specified  until Secton 3. We will 
only assume that the protons form a continuum of single 
particle states and that they are in the mixed type II 
superconducting state.
As an outer interface bounding the neutron-proton superconducting phase
we shall consider the crust-core interface, the location of which is 
identified with either the density of transition of the protons 
from a continuum to clustered state or the vanishing of neutron/proton 
superfluidity with decreasing density. For definiteness we shall 
proceed with the first condition, i.e. assume that the external interface 
is located at the position where the protons become unstable against 
clustering into nuclei. In that case the transition is first-order,
and the interface is of the order of one nuclear spacing thick,
in analogy with an ordinary liquid-solid interface (Pethick, 
Ravenhall \& Lorenz 1995).
The location of the inner core-outer core interface 
will be identified with the disappearance of the superconducting 
state of neutrons or protons with increasing density. 
For brevity we shall sometimes refer to these interfaces 
as the external and internal interfaces, respectively. The
Bose condensation of mesons  provides another possibility, that the 
inner interface is located at the density of the onset of the 
developed condensate.

\subsection{Vortex - external interface interactions: cylindrical
geometry}

To set up the problem, assume that the crust supports a certain magnetic 
field of strength $H_0$: a plausible origin for such a field could be
the  thermal battery effect (Blandford, Applegate \& Hernquist 1983, 
Urpin, Levshakov \& Yakovlev 1986) or the natal dynamo effect (Duncan 
\& Thompson 1993).
The type II superconducting protons in the core are assumed to be in the 
mixed state. The magnetic field, $\bmath B_v$, of quantum vortices in 
the superfluid core is governed by the London equation for 
charged-neutral superfluid mixtures (Vardanian \& Sedrakian 1981;
Alpar, Langer \& Sauls 1984; Sauls 1989):
\be\label{LONDON} 
\delta_p^{2}\,\bmath\nabla &\times& (\bmath\nabla \times \bmath B_v)
+\bmath B_v =
\sum_{\tau} \bnu_{\tau}\, \, \Phi_{\tau}\nonumber \\
&\times& \sum_q\left[\delta^{(2)}
\left(\bmath r-\bmath r_{\tau \, q}^{(+)}\right)-
\delta^{(2)}\left(\bmath r-\bmath r_{\tau \, q}^{(-)}\right)\right],
\ee
where $\delta_p$ is the magnetic field penetration depth;
the $\bnu$'s are circulation unit vectors; the $q$-summation 
is over the vortex sites in the two-dimensional vortex lattice plane;
$\tau = \pm 1/2$ sums over the isospin
projection; $\Phi_{1/2} \equiv \Phi_0$ is the flux quantum carried 
by proton vortices and  $\Phi_{-1/2} \equiv k \Phi_0$ is the non-quantized 
flux of neutron  vortices due to the entrainment effect; 
the entrainment coefficient $k$ is a continuous function of 
nucleon effective masses
(Andreev \& Bashkin 1976;
Vardanian \& Sedrakian 1981; Alpar, Langer \& Sauls 1984; Sauls 1989;
the Fermi-liquid corrections to the entrainment coefficient
have recently been discussed  
by  Borumand, Joynt \& Kluzniak 1996). 
The second term on the right-hand side of equation (\ref{LONDON})
takes into account the attractive part of the vortex-interface interaction 
in terms of vortex images of opposite sign, which are  
located symmetrically  with respect to the crust-core interface. 
The total field at the crust-core interface, $\bmath B$, is the 
superposition of the solution of equation (\ref{LONDON}),
\be\label{BVORTEX}
\bmath B_v &=& \sum_{\tau}\bmath\nu_{\tau} \frac{\Phi_{\tau}}{2\pi\delta_p^2}
\nonumber \\
&\times&\sum_{q}\left[K_0\left(\frac{\vert \bmath r-\bmath r_{\tau\,q}^{(+)}\vert}{\delta_p}
\right) - K_0\left(\frac{\vert \bmath r -\bmath 
r^{(-)}_{\tau\,q}\vert}{\delta_p}\right)\right], 
\ee
and the induction field $\bmath B_{cr}$ set up by the 
crustal field $\bmath H_0$, which exponentially 
penetrates in the superfluid core to a scale of the order 
of $\delta_p$; (here and below the $K$ are the modified Bessel functions.)
The magnetic field distribution with the proper boundary 
condition at the interface
allows one to calculate the relevant part of the Gibbs free
energy of the system $G = F - (4\pi)^{-1}\int \bmath B\cdot \bmath H_0 dV$, 
where  the free-energy is  
\be\label{GIBBS} 
F&=& 
\frac{\delta^2_p}{8\pi}\int\left[\,\bmath B\times 
(\bmath \nabla \times \bmath B) \,\right]\cdot d\bmath S \nonumber \\
&+&\frac{\delta^2_p}{8\pi}
\int\!\bmath B\cdot[~\delta_p^{-2}\bmath B+\bmath\nabla\times(\bmath \nabla 
\times \bmath B)~]~dV .
\ee
The surface integration is over the crust-core interface,
while the bulk integration involves the superfluid core region. Assume 
the vortex lattice plane is the $(xy)$-plane of Cartesian system of 
coordinates, with the interface being the  $(yz)$-plane, and the circulation
vectors in the superfluid core region are in the positive $z$-direction;
the half-plane $x<0$ corresponds to the crust 
while $x>0$ corresponds to the superfluid core; see Fig. 1.
\begin{figure}
\begin{center}
\mbox{\psfig{figure=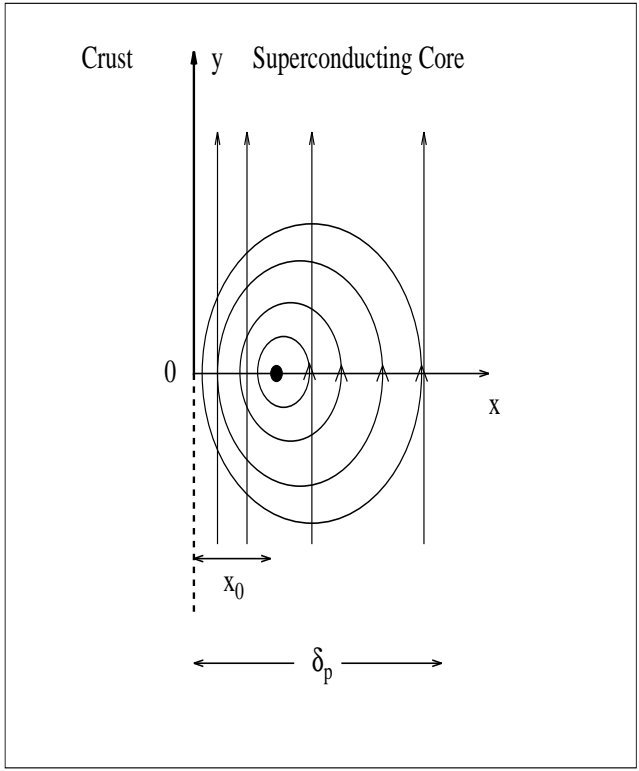,height=3.3in,width=4.in,angle=-90}}
\end{center}
\caption[] 
{\footnotesize{ The geometry of vortex - crust-core interface interaction. 
The circles are the  streamlines of the proton supercurrent circulation, 
which is screened beyond the scale $\delta_p$. 
The Meissner currents (straight lines) induced by the 
crustal magnetic field  penetrate within the superconducting core on 
the scale $\delta_p$.
The repulsive force arises because of the interference of oppositely directed
Meissner currents and the vortex circulation. The attractive term arises
because of  the 
deformation of the vortex circulation near the boundary, where the density of 
the streamlines must be higher and, therefore, the superflow velocity is 
larger than on the opposite side of the vortex.
This deformation is accounted for by adding the image of the vortex, 
as discussed in the text.
The velocity gradients translate into pressure gradients,
resulting in the vortex-interface interaction force.}}
\label{fig1}
\end{figure}
\begin{figure}
\begin{center}
\mbox{\psfig{figure=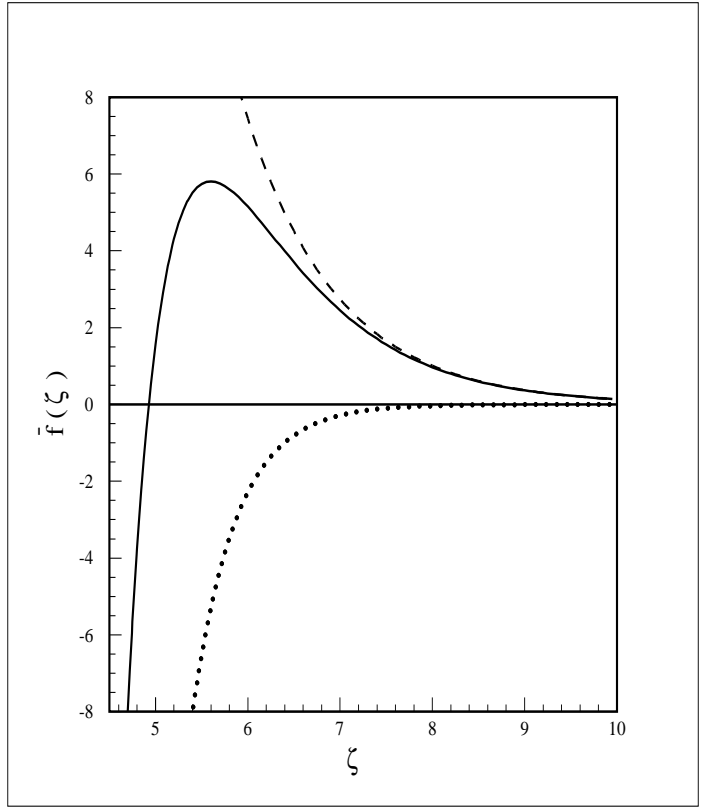,height=4.2in,width=3.2in,angle=0}}
\end{center}
\caption[] 
{\footnotesize{The dimensionless vortex-interface interaction force 
$\overline  f(\zeta)\times 10^{-6}$ as
a function of the reduced distance $\zeta$ for the crustal magnetic 
field value $H_0 = 10^{12}$ G  ({\it full line}). The {\it dashed line} 
shows the repulsive component, while the {\it dotted line} is the 
attractive component of the force. For a discussion of the asymptotic 
behavior see the text. }}
\label{fig2}
\end{figure}
The interactions, therefore, are $x$-dependent; the crustal magnetic field has 
a $z$-component of the form $B_{cr}= H_0\, {\rm exp}\left(-x/\delta_p\right)$, 
and the boundary condition for the total magnetic field is $B_z = H_0$.
Integrating equation (\ref{GIBBS}), one finds $G = \sum_{\tau q}  
G_{\tau q} (x^{(+)})$, where 
\be\label{GIBBS_2}
G_{\tau \, q}(x^{(+)})&=& \frac{\Phi_{\tau}}{4\pi}
\Biggl\{H_0\, {\rm exp}\left(-\frac{x_{\tau\, q}^{(+)}}{\delta_p}\right) 
-H_0 \nonumber \\
&+&\frac{\Phi_{\tau}}{4\pi\delta^2_p} 
\left[{\rm ln}\left( \frac{\delta_p}{\xi}\right)-
K_0\left(\frac{2\, x_{\tau\, q}^{(+)}}{\delta_p}\right)\right]\nonumber \\
&+&\sum_{\tau'}\frac{\Phi_{\tau'}}{4\pi\delta^2_p}
\sum_{q'}^{\prime}\Biggr[K_0\left(\frac{\vert x_{\tau'\, q'}^{(+)}
-x_{\tau\, q}^{(+)}\vert}{\delta_p}\right)\nonumber \\
&-&K_0\left(\frac{\vert x_{\tau'\, q'}^{(+)}
- x_{\tau\, q}^{(-)}\vert}{\delta_p}\right)\Biggl]\Biggr\},
\ee
$x^{(+)}$ and $x^{(-)}$ denote the positions of the vortex and the 
image, respectively, and $\xi$ is the coherence length of proton 
superconductor.  Note that in the self-energy term we introduced  a cut-off
$K_0(x) \simeq {\rm ln}(\delta_p/\xi_p )$ for $x\to 0$. The primed summation
assumes that the terms $q=q'$ and $\tau =\tau'$ are omitted.

The force associated with the interaction part of the 
Gibbs potential (\ref{GIBBS_2}) is 
\be\label{FORCE} 
f_{\tau \, q}(x^{(+)})&=&\frac{\Phi_{\tau}}{4\pi}\Biggl\{\Biggl[\frac{H_0}{\delta_p}\,
 {\rm exp}\left(-\frac{x_{\tau\, q}^{(+)}}{\delta_p}\right) -\frac{\Phi_{\tau}}{4\pi\delta^3_p} 
K_1\left(\frac{2\, x_{\tau\, q}^{(+)}}{\delta_p}\right)\Biggr]\nonumber \\
&+&\sum_{\tau'}\frac{\Phi_{\tau'}}{4\pi\delta^3_p}
\sum_{q'}^{\prime}\Biggl[K_1\left(\frac{\vert x_{\tau'\, q'}^{(+)}
-x_{\tau\, q}^{(+)}\vert}{\delta_p}\right)\nonumber \\
&-&K_1\left(\frac{\vert x_{\tau'\, q'}^{(+)}
- x_{\tau\, q}^{(-)}}{\delta_p}\right)\Biggr]\Biggr\}.
\ee
The first term in equation (\ref{FORCE}) corresponds to the 
repulsive force acting between the vortex magnetic flux  and the crustal 
magnetic field. It can also be interpreted as a Lorentz force resulting 
from superposition of velocity fields of the vortex and the surface Meissner 
currents (Fig. 1). The second term is the force between 
the vortex and its image, which is attractive because of oppositely 
directed circulation vectors. The last two terms correspond to vortex-vortex
and vortex-image interactions, respectively. 

\subsection{Single Vortex - Interface Interaction}

To estimate the magnitude and asymptotics of the vortex-interface interaction,
consider a single  vortex with  flux $\Phi_0$. 
From equation  (\ref{FORCE}) we find 
\be\label{REDUCED_FORCE}
\overline f(\zeta)= \left[\overline H\, e^{-\zeta} - K_1(2\zeta)\right],
\ee
where the dimensionless variables are defined 
as $\overline f(\zeta) = f(x)/f^*$, 
$f^* = \Phi_0^2/8\pi^2\delta_p^3$,
$\overline H = H_0/H^*$, $H^*=\Phi_0/2\pi\delta_p^2$, 
and $\zeta = x/\delta_p$. The function $\overline f(\zeta)$ is plotted
in Fig. 2  for $\overline H = 0.003$. For large distances 
($\zeta \to \infty$) the exponential (repulsive) term dominates,
because in this limit the second term  goes to zero more rapidly, 
[$K_1(2\zeta)\propto \sqrt{\pi/4\zeta} ~e^{-2\zeta}$].
For small distances the second (attractive) term in 
equation (\ref{FORCE}) dominates; note that it should 
be cut-off at $\zeta\sim \xi/\delta_p$ 
by replacing $K_1(2\xi/\delta_p)$ by $\delta_p/ 2\xi$.
The repulsive part of the vortex - crust-core interface 
interaction, which dominates at large distances acts as a potential 
barrier to a vortex approaching the boundary, and thus  prevents
its continuous decay on the interface. If the 
magnitude of the crustal magnetic field is decreased, the potential barrier 
disappears  in the limit $\overline H\to 0$. 

The crust-core interface density will be identified with the 
density of separation of two phases where the protons are 
clustered in the nuclei (crust) and are in the continuum state (core).
Following Pethick, Ravenhall \& Lorenz (1995) we adopt the phase transition
density $\rho_{\rm tr}\simeq 1.56 \times 10^{14}$ g cm$^{-3}$. 
As pointed out earlier, the phase transition found in Pethick,
Ravenhall \& Lorenz (1995)
is an analog of the ordinary first-order liquid-solid phase transition 
and the interface between the phases is one nuclear spacing thick. 
The stability of the interface requires a positive curvature energy 
associate with the interface between two phases; the system, therefore, 
will tend to minimize the surface of the interface. 
The typical scale of non-homogeneity should be the largest of the two 
relevant scales in the problem - the actual size of the nuclei and  
the size of the nuclear spacing. Both length scales are of the 
order of 30-50 fm. On the other hand, the interaction range, as can be seen
from Fig. 2, is on order of 5-10 $\delta_p$ which translates 
into 1-5 $\times 10^3$ fm for the value  $\delta_p \simeq 200$ fm 
found at  $\rho_{\rm tr}$  (see Table 2 later). As the interaction range
exceeds the thickness of the interface by two orders of magnitude, 
the finite size of the interface can be ignored in the 
calculation of the force acting a vortex, which justifies the treatment 
above.
Using the same transition density as above, we find
$H^* \simeq 3.3 \times  10^{14}$ G, $f^* 
\simeq 5.4\times 10^{17}$ dyn cm$^{-1}$
for $\delta_p= 100$ fm.
Assuming a conventional value for the crustal magnetic field 
$H_0 = 10^{12}$ G, we obtain $H_0/H^* = 0.003$ and the maximum value of the 
reduced force $\overline f^{\rm max} \simeq 5.8 \times 10^{-6}$ at
$\zeta =  5.6 $. This translates to the maximal repulsive force   
$f^{\rm max} =f^*~\overline f^{\rm max}\simeq 3.13 \times 10^{12}$ 
dyn cm$^{-1}$. In general, the magnitude of the crustal magnetic 
field at the  crust-core interface for different objects 
can vary in a reasonable range, $10^9< H_0<10^{13}$ G, though 
values beyond both extremes cannot be excluded. 
For further reference the values of the maximal force
for different applied magnetic fields are given in Table 1. 
In the range of interest, the maximal force depends  quadratically on 
magnitude of the crustal magnetic field.
\begin{table}
\begin{center}
\caption{Values of the maximal force for different applied magnetic fields.}
\begin{tabular}{ccccc}
\hline
${\rm log}(H_0)$ [ G ]& $\overline f^{\rm max}(\zeta_0)$ [dyn/cm] 
& $\zeta_0$\\
\hline
${13}$ & $4.5\times 10^{-4}$  & 3.6 \\
${12}$ & $5.8\times 10^{-6}$  & 5.6 \\
${11}$ & $7.0\times 10^{-8}$  & 7.6 \\
${10}$ & $7.8\times 10^{-10}$  & 9.8 \\
${9}$  & $8.7\times 10^{-12}$  & 12.1 \\
\hline
\end{tabular}
\end{center}
\end{table}

\subsection{Vortex - external interface interactions: spherical
geometry}

The vortex-interface interaction in its simplest form, 
equation (\ref{REDUCED_FORCE}), is easily generalized to the case where
the vortex and the crustal magnetic field from an arbitrary angle
with the spherical interface. Assume a straight vortex segment along 
the $z$ direction of the cylindrical-polar coordinates. For a spherical 
interface of radius $R$ we find:
\be 
\overline f(\varrho, \theta) &=& {\rm Sin}\,\theta\,\left[\overline H_z\, 
e^{-\zeta}-K_1\left(2\zeta\right)\right], \nonumber\\ 
\zeta(\varrho ,\theta) &=& \delta_p^{-1}
\left(R-\rho {\rm Sin}^{-1}\theta\right), 
\ee
where $\varrho$ is the distance of the vortex segment from the 
rotation axis in cylindrical coordinates with the origin at the 
$z$-axis, $\theta$ is the polar angle, $\overline H_z$ is the 
projection of the crustal field on the $z$-axis. Because of the 
weak angle dependence of the force, our previous order of magnitude 
estimates  remain valid unless the angle $\theta$ is 
close to zero, (i.e. the segment is close to the rotation axis). 
As the force vanishes exponentially for large arguments, it 
acts only at the ends of a vortex enclosed in a spherical shell. If the 
vortex is assumed to be straight, the averaging of the vortex - interface 
force over its length reduces the effective net force by the ratio of 
the effective range of the interaction to the length of the vortex. 
The vortices participating in the glitch generation process need to 
adjust the (spherical) configuration of the interface in order to sustain 
the required magnitude of the Magnus force in the interjump period. 
In the spherical geometry this implies that the vortices, the ends of which 
are fixed at the interface, bend, forming an arc with a curvature radius of
the order of $R$. The balance between the local tension force and the 
vortex-interface force determines whether the vortices in the bulk
of the fluid will in fact bend under the action of the Magnus force. 
The magnitude of the force resulting from the vortex tension is given via the 
vortex self-energy as
\be 
f_T = \left(\frac{\Phi_0}{4\pi\delta_p}\right)^2{\rm ln}
\left(\frac{\delta_p}{\xi}\right) {\cal R}^{-1}, 
\ee
where ${\cal R}$ is the radius of the vortex curvature, with  
maximal value for the arc of the order of $R$.
Assuming $\delta_p/\xi_p = 10$, $f_T = 5\times 10^{6}$
dyn $(R/{\rm cm})^{-1}$, which is much smaller than the 
local vortex - interface interaction force. Thus, while the 
ends of the vortex are fixed at the interface, the Magnus 
force will bend the vortex in the bulk of the superfluid.
Eventually the vortex will assume the form of the interface
and the vortex-interface force will balance the Magnus 
force along the entire length of the vortex. Note that
there is a cumulative increase of the vortex-interface force 
with the bending of the vortex - the larger the vortex curvature 
the larger is the net effective force. As the quantitative analysis
now can be carried out using the local form of the force balance 
equation, in complete analogy to the previous discussion, we shall 
not repeat it here. 
  
\subsection{Vortex-internal interface interactions}

The main difference between the physics of interaction of the vortices 
with the external and internal interfaces is that in the latter case 
the flux in the normal region enclosed in a superconductor 
must be quantized in units  of $\Phi_0$ (the superconducting 
region is multi-connected). For simplicity we shall 
consider a cylindrical geometry where external and internal 
interfaces are coaxial cylinders with the axis along the 
vector of rotation. The equilibrium value of the magnetic field 
in the inner normal core is established from a detailed balance 
between  the processes of quantum flux capture from the superconducting 
region and flux drift in the superconducting region. 
In equilibrium, the number of quantum fluxes trapped in the 
internal region should minimize the energy of the system. 
The spin-down of the star
or the Ohmic decay of the field within the normal core will drive
the system out of equilibrium;
e.g. if the magnetic field decreases in the 
inner core because of  Ohmic dissipation, 
the flux transport into it would affect the 
dynamics of the neutron lattice  requiring contractions, i.e. spin-ups in 
the angular velocity of the superfluid and, respectively,
slow-downs in that of the crust.
Conversely, spin-down would require formation of new vortices 
at the interface and their entrance into the superconducting region. 
The barrier at the inner core-outer core interface 
would prevent this processes from being continuous.

Let us derive the expression for the barrier at the inner interface. 
The magnetic field in the superconducting region 
bounded by two interfaces ($r_{\rm in}\le r\le r_{\rm out}$)
is determined by the London equation (\ref{LONDON})
with the boundary conditions
\be 
B(r_1)= H_*, \quad B (r_2)= H_0,
\ee
where  $H_*$ is the magnetic field strength within the 
inner core. 
The solution of the homogeneous London equation is 
\be\label{BIN}
B(r) = \frac{I_0(r)}{I_0(r_2)} H_0 \left[1 + \gamma(r) \right] 
-\gamma (r) {\cal H},
\ee
where 
\be\label{gamma} 
\gamma (r) = \frac{K_0(r)I_0(r_2)-K_0(r_2)I_0(r)}
{K_0(r_2)I_2(r_1) - K_2(r_1) I_0(r_2)} 
\ee
and $K_n$ and $I_n$ are modified Bessel functions. Here the field 
${\cal H}$ is the sum of the quantized field trapped in the inner core
and the field generated by the entrainment currents at the inner interface,
\be 
{\cal H} &=& \frac{q\Phi_0}{\pi r_1^2} 
+ \frac{ 2 m c \vert k\vert }{e} \Omega,
\ee
where $q$ is an integer.
The result (\ref{BIN}) is derived by combining  the equation for
the circulation of the vector potential,  $\bmath A$, at the boundary
of the inner core, 
\be 
\oint \bmath A\cdot \bmath l = \pi r_{1}^2 H_*
\ee 
with the Maxwell equation for the magnetic field in the 
form 
\be 
\delta_p^2\left(\bmath\nabla \times \bmath B 
-\frac{4\pi}{c}\bmath j_{12}\right)= 
\frac{\Phi_0}{2\pi}\bmath\nabla\chi_1-\bmath A , 
\ee
where $\chi_1$ is the phase of the proton superconductor and 
$\bmath j_{12} = \rho_{12}\bmath v_{2}$ 
is the entrainment current (see Appendix~\footnote 
{Here and in the Appendix we use unconventional, but notationally
convenient, isospin indices 1 and 2 for protons and neutrons respectively.}).
For the geometry adopted (with cylindrical coordinates $r,\phi,z$),
the functions have  simple azimuthal
dependences; in particular the phase of proton supercurrent 
along the boundary is quantized:
\be 
\left(\bmath \nabla \chi \right)_{\phi}  = \frac{q}{r}; 
\ee
and 
\be
\left( \bmath\nabla \times \bmath B \right)_{\phi} 
= -\frac{\partial B_z}{\partial z}; \quad
\left( \bmath j_{12}\right)_{\phi}  = \frac{e}{m}\rho_{12}\Omega r,
\ee
where $\Omega$ is the rotation frequency and $\rho_{12}$ is the 
density of the entrained protons.  The total field 
is given by the sum of equation (\ref{gamma}) and the particular solution 
equation (\ref{BVORTEX}) (owing to the short range of the spreading of the 
vortex field around the singularity, the vortex contribution to the 
boundary condition can be omitted). Knowledge of the total magnetic 
field allows one to calculate the Gibbs potential as before. One finds 
$G = \sum_{\tau q}  G_{\tau q} (x^{(+)})$, where 
\be\label{GIBBS_3}
G_{\tau \, q}(x^{(+)})&=& \frac{\Phi_{\tau}}{4\pi}
\Biggl\{  B\left( \frac{x_{\tau\, q}^{(+)}}{\delta_p} \right)
-{\cal H} \nonumber \\
&+&\frac{\Phi_{\tau}}{4\pi\delta^2_p} 
\Biggl[{\rm ln}\left( \frac{\delta_p}{\xi}\right)
-K_0\left(\frac{2\, x_{\tau\, q}^{(+)}}{\delta_p}\right)\Biggr]\nonumber \\
&+&\sum_{\tau'}\frac{\Phi_{\tau'}}{4\pi\delta^2_p}
\sum_{q'}^{\prime}\Biggr[K_0\left(\frac{\vert x_{\tau'\, q'}^{(+)}
-x_{\tau\, q}^{(+)}\vert}{\delta_p}\right)\nonumber \\
&-& K_0\left(\frac{\vert x_{\tau'\, q'}^{(+)}
- x_{\tau\, q}^{(-)}\vert}{\delta_p}\right)\Biggl]\Biggr\}.
\ee
It can be seen that this result differs from that for the 
external interface in the functional form of the field at the 
interface. In deriving this result it has been assumed that
the curvature of the inner surface is much larger than the 
size of a proton vortex and, therefore, 
the interface curvature can be ignored. The maximal force is 
given, as before, by the maximal value of the derivative of 
equation  (\ref{GIBBS_3}). The existence of the inner 
interface is a speculative issue because of our limited knowledge of 
the state of matter beyond several times the nuclear saturation 
density and the nature of pairing at these densities. 
We shall not evaluate quantitatively the results of this subsection.
One may  note, however, that if mesons form a 
Bose condensate, the interface between
the proton superconductor and the pion/kaon condensate
would have a barrier that would be a function of the 
differences in the magnetic field penetration depths of the 
condensates.

\section{Collective Interactions}
\subsection{Vortex Clusters}

For further progress, we needs to specify the ground state structure 
of vortices in the superfluid core. The 
proton vortices would mediate the interaction
between the neutron vortices and the interface on quite 
general grounds because, first, the neutron and proton vortex 
interactions with the interface
are of the same order of magnitude, and second, the total 
number of proton vortices per area of a neutron vortex is very large:
$n_1 = \left({B}/{\Phi_0}\right)\,\left({\kappa}/{2\Omega}\right)
= 2.5 \times 10^{13}~B_{12}~ \Omega^{-1}_{190}$,
where $\Phi_0 = \pi\hbar c/e$, $\kappa = \pi\hbar/m$ ($m$ being
the neutron mass) $B = 10^{12}B_{12}$ G is the mean magnetic field
strength and $\Omega = 190 \, \Omega_{190} $ s$^{-1}$ is the spin
frequency of the Crab pulsar.

We shall adopt here the vortex cluster model in which 
the distribution of $10^{13}$ proton vortex lines per neutron vortex 
is inhomogeneous and these  
are confined in a cluster around the neutron vortex (see Papers
I and II). 
The scenario for vortex nucleation in the presence 
of the primordial magnetic field 
of a neutron star has been suggested by Baym, Pethick \& Pines (1969); 
(see also Jones 1987 and references therein 
for the dynamics of uncoupled proton vortices in the core). 
They  show that the Meissner state is preferable for fields 
lower than the lower 
critical field  $H_{c1}\sim 10^{14}$ G; however, 
it cannot be achieved because the 
flux expulsion time is comparable to the lifetime of the pulsar. 
Therefore the field is forced to nucleate in the superconducting state
via a {\it first}-order phase transition in a mixed state, even when this 
state is not the thermodynamically favoured one. 

We suggest here a modified scenario which takes the advantage of 
the phase transition being of the first order.
The kinetics of three-dimensional nucleation would initially
require  a field expulsion from randomly localized seeds
of the superconducting state. The first-order phase transition  would be 
realized  via creation and subsequent expansion of the seeds of the 
stable (superconducting)  phase within the metastable (normal) state.
The nucleation process goes through two stages: first, formation of 
superconducting seeds of the critical size, 
which are, thus, stable against the collapse
back into the normal state; and secondly, 
coalescence of supercritical seeds
of the superconducting phase.
This process would lead to a squeezing of the field into  normal domains
on  microscales; when the size of the seeds of the 
superconducting phase becomes large enough, the field in the normal domains 
will exceed $H_{c1}$. After this the nucleation of proton vortices 
will become energetically favorable. 
As the magnetic flux scales as the 
square of the linear size of a region, by flux conservation,
the squeezing of an initially homogeneous field $\sim 10^{12}$ G  
over the area of a neutron vortex 
to a scale one order of magnitude smaller will 
drive the field intensity beyond the critical value $\sim 10^{14}$ G.
The proton vortices will be arranged in clusters with linear 
size $\sim 0.1 d_n$ (where $d_n$ is the neutron inter-vortex spacing)
and will sustain a mean magnetic field induction  $\ge H_{c1}$.
For fields higher than  $H_{c1}$, the vortex state 
will nucleate in a homogeneous vortex structure.

Thermodynamic considerations show that intrinsic nucleation 
of clusters in response to the 
superfluid-dynamo is another possible mechanism; in this case,
however, the structure and arrangement of the clusters can 
be predicted without resorting to kinetic theory of nucleation, 
and these are confined around the neutron vortex and are parallel 
to the neutron vortex lattice (see Papers I, II, and 
Appendix A of this paper). 

What is the vortex cluster configuration when both 
creation mechanisms - those resulting from the residual field and 
the superfluid dynamo effect  - are present? 
The answer perhaps depends on the sequence in which 
the neutron superfluidity and the proton superconductivity set in. 
The superfluid gap density profiles in the core 
show that the protons condense first; a naive expectation
would be that the superfluid dynamo (present only if 
neutrons are superfluid) should operate in the presence of 
the proton vortex state because of the residual field. This might
not be true because, on one hand, the first order transition to 
the superconducting state is
commonly strongly delayed, so that the transition occurs from an
 overcooled state,  and on the other hand,
the pulsar temperature drops quickly below 
the critical temperatures of phase transitions.
If both structures nucleate {\it independently}, their dynamics would 
be coupled via cluster-cluster interactions. An expression for the 
interaction force is given by the fourth line of eq. (\ref{FORCE}),
\be\label{FORCE_CLUSTER} 
\vert\bmath f_{\rm int}\vert = 
\sum_{\tau \tau' q q'}^{\prime} \bmath\nu_{\tau}\cdot
\bmath\nu_{\tau'}
\frac{\Phi_{\tau}\Phi_{\tau'}}{16\pi^2\delta^3_p}
\Biggl[K_1\left(\frac{\vert x_{\tau'\, q'}^{(+)}
-x_{\tau\, q}^{(+)}\vert}{\delta_p}\right)\Biggr],
\ee
except that in general it depends on the angle formed by the circulation 
vectors of the clusters $\bmath \nu_{\tau}$. If this 
is less than a right angle the cluster would
repel each other; 
in the opposite case these would merge and eventually annihilate.
Note that the main contribution to this interaction comes from the  
changes in the {\it kinetic energy } of the clusters. A
similar expression has been derived by Srinivasan et al. (1990),
who recognize it as a repulsive force, Jones (1991b),
Ruderman (1991), Mendell (1991) and Ding, Cheng \& Chau (1993).
The force is sometimes termed as a pinning interaction.
Consistent with our treatment of the vortices as singular 
lines, the force does not include the modifications of the 
core energy of the vortex lines (Sauls 1989).
Because of the large number of proton vortices collected in a cluster 
($\sim 10^{13}$), the mutual creep considered by Ding, Cheng \& Chau  (1993)
when the vortices form a homogeneous array,  would be prohibited in our case
by the large potential barrier of the vortex cluster.

While the problem of field nucleation and cluster-cluster interactions in the 
bulk of the superfluid needs more detailed consideration, 
we shall restrict ourselves in the following material to clusters created 
by the superfluid dynamo effect. We believe that our results 
would not change qualitatively in a more elaborate picture.

\subsection{Estimates}

The localization radius of clusters, independent of the details of their
nucleation mechanism, is
of the order $\delta_n \sim 0.1 ~d_n$, where $d_n$ 
is the neutron inter-vortex spacing 
(unless the initial field is larger than the lower critical one). 
As the proton intervortex distance 
$d_p \ge 10~\delta_p$ and the range of interaction 
is several times $\delta_p$, a single row of 
vortices will interact with the interface.
The number of vortices in a row is 
$ N_{\Phi}\simeq\sqrt{8\,\delta_n\,\delta_p/d_p^2}$.
\footnote{
This estimate follows from the observation that the length of a 
row lying within the range of the interaction from cluster's 
circular  boundary 
is $2\sqrt{\delta_n^2-(\delta_n-\Lambda)^2} = 2 \sqrt{2\delta_n\Lambda+
O(\Lambda^2)}$, where $\delta_n$ is the cluster radius, 
$\Lambda\sim\delta_p$ is the effective range of the 
interaction. The number of vortices is obtained by dividing the length
of the row by the lattice constant $\sim d_p$.}

Using the value $d_p\ge 10^3$ fm  for a Vela-type pulsar,
one finds $N_{\Phi} \simeq 283$.
The maximal force on the vortex cluster, therefore, 
is $f^{\rm max}_C = N_{\Phi}~ f^{\rm max} = 8.9\times 10^{14}$ dyn cm$^{-1}$.
The repulsive component of the vortex-interface interaction force 
would require an increase in the Magnus force by a magnitude    
\be\label{MAGNUS}
f^M &=& 3.23\times 10^{17}
\left(\frac{\delta\omega_s}{s^{-1}}\right)\nonumber \\
&\times&\left(\frac{\nu}{1.98\times 10^{-3}~{\rm ~cm^{2}~s^{-1}}}\right)\,
\left(\frac{R}{9.6\times10^{5}~{\rm cm}}\right)\,
\ee
per neutron vortex, in order to restore the free-flow expansion of the 
neutron vortex lattice through the interface. Here 
$\rho_s$ is the superfluid density, $\nu = \pi\hbar/m_n$, 
$m_n$ is the neutron mass, $R$  is the distance of the crust-core 
boundary from the rotation axis, and $\delta\omega_s$ the 
angular velocity departure between the superfluid and the 
normal components.\footnote{The neutron star model used in 
our estimates is discussed in detail in Paper II.} 
From the  balance condition $f^{\rm max}_C = f^M$ one finds 
the value of the maximal departure $\delta\omega_s^{\rm max} \simeq 0.003$ 
s$^{-1}$ that can be sustained  by the boundary force on the cluster. 
Note that although the vortex-interface force is effectively acting on 
the outmost neutron vortex dressed by a proton vortex cluster, the long-range
neutron vortex-vortex interaction ensures that the force is acting on 
a macroscopic domain of neutron vortex lattice.

One can now estimate whether the effective magnitude of 
the moment of inertia of the superfluid vortex domain near the 
crust-core interface, which has short $(\le 120$ s) dynamical coupling 
times, is sufficient to drive a large Vela-type glitch, provided 
the angular velocity departure has reached its critical
value $\delta\omega^{\rm max}_s$ . Note that, although 
the range of the interaction is microscopic, a macroscopic 
region of the superfluid neutron vortex lattice 
will undergo compression because of stopping of the 
continuous vortex current through the interface. 
For large Vela glitches,
the magnitude of the jump in the rotation rate of the normal 
component is $\delta\omega/\omega \sim 10^{-6}$. Using the 
angular momentum conservation $I_s\delta\omega_s^{\rm max} 
= I_n \delta\omega$, where $I_s$ and $I_n$ are the 
moment of inertia of the superfluid and normal components, 
respectively, for the Vela pulsar we find $I_s/I_n = 0.023$.
For a comparison with the model calculations one needs to know 
the  moment of inertia of the normal component which undergoes the 
observed spin-up. 
For the neutron star model described in Paper I,
we find $I_s/I_n = 0.017$ assuming that the $I_n$ is the whole crust.
This number is  close to the estimate above, but is uncertain because
of our limited knowledge of the number of components of the star involved 
in the short spin-up process.
   
\section{Spin-up Timescales}

In discussing the spin-up time-scales we shall assume that 
the spin-up of the normal component (or at least the layers relevant 
for generation of the glitch) can be separated from the 
spin-up of the superfluid due to the mutual friction. Our conclusions
can be modified in a combined treatment of  the normal and superfluid 
components, because the spin-up times for the core plasma 
(in the Vela pulsar) are of the same order of magnitude 
as that for the superfluid. Easson (1979) has discussed the Ekman pumping
mechanism of the normal component and has found spin-up time-scales on the 
order of 10 s. Stratification effects (Goldreich \& Reisenegger 1992)
change this picture; in particular the  Ekman pumping cannot explain the 
fast core-crust coupling in the Vela pulsar (Abney, Epstein \& Olinto 1996).  
Lee (1995) and Mendell (1998) discussed the coupling via the superfluid 
oscillations;  the cyclotron-vortex waves were identified as a
being important ingredients of the spin-up process (Mendell 1998).

Let us first  consider the dynamics of the electron liquid and separate
the volume per single neutron vortex in the region including the proton
vortex cluster $V_<$ and the remaining volume, $V_>$,
which is free of magnetic flux. The collision integral in the Landau-Boltzmann 
equation for the electron distribution receives its two main contributions 
from electron-electron ($e$-$e$) and electron--proton vortex ($e$-$\Phi$) 
collisions. The rate of the first process is given by the reciprocal of the  
lifetime of a  quasiparticle in a relativistic electron
Fermi-liquid, $\tau_{ee}$, as 
\be\label{TAUEE} 
\tau_{ee}^{-1} = \frac{\pi^3}{16} \frac{(k_BT)^2}{\hbar \epsilon_{eF}} F(\beta),
\ee  
where $T$ is the temperature, $k_B$ is the 
Boltzmann constant,  $\epsilon_{eF} = \hbar c k_{Fe}$ 
is the Fermi energy of electrons, $k_{Fe}$ is
the electron Fermi wave number, and 
\be 
F(\beta) &=& \frac{\beta}{1+\beta}+\beta^{1/2}{\rm sin}^{-1}(1+\beta)^{-1/2}
\nonumber \\
&-&
\frac{\beta}{(1+2\beta)^{1/2}}{\rm cos}^{-1}\left(\frac{\beta}{1+\beta}\right)
\ee
is a correction arising due to the finite range of the interaction, where 
$\beta = k_{TF}^2/4 k_{Fe}^2$ and $ k_{TF}$ is the Thomas-Fermi 
screening length of  
protons (Smith \& Hojgaard Jensen 1989; for the screening effects see  Baym et al 1969). 
The rate for the second process, electron-flux scattering, 
in the volume $V_<$ is 
\be 
\tau_<^{-1} = \frac{3\pi^3}{64}\frac{c\delta_p}
{(k_e\delta_p)^2}~\frac{\overline B}{\Phi_0},
\ee 
where $\overline B\simeq 10^{14}$ G is the mean magnetic field of the cluster. 
In the volume $V_>$ the $e$-$e$ rates are given by the same expression
(\ref{TAUEE}), however the electron flux 
scattering is only on the boundaries of the clusters. The rate of this 
process can be estimated as
\be\label{LIMIT} 
\tau_>^{-1} \simeq
n_n \, c \, d_n\,  \left(\frac{\xi_p}{\delta_p}\right)^{1/2\vert k\vert},
\ee
where $n_n$ is the neutron vortex number density, which is related 
to the neutron vortex (triangular) lattice constant, $d_n$, by the relation 
$n_n = 2/\sqrt{3} \, d_n^2$. 
This estimate does not take into account the finite 
`skin' of the cluster and  a more precise calculation
would require inclusion of 
the structure factor of the vortex cluster, 
which would lead to somewhat larger scattering rates. The estimate 
(\ref{LIMIT}), which assumes an impenetrable cluster, 
is the opposite limit of the relaxation time found 
in Paper I, which assumes a transparent cluster. These time-scales give 
the upper and lower bounds on the relaxation times, although the first
one should be closer to the exact value. The ratio of these time-scales 
is of order of unity in the density region of interest, but can increase
by two orders of magnitude at high densities.
The values of relaxation times
along with the relevant microscopic parameters (Baldo et al 1992)
are given in Table 2.

The last two columns give 
the corresponding viscosity and the dynamical spin-up time scale. 
It can be seen that in the region $V_<$ 
one has $\tau_{ee}\gg\tau_<$, which means that 
electrons are localized within the clusters (i.e. the electrons equilibrate 
among themselves much slower than with the vortex lattice).
In this case the electron fluid can not be 
ascribed an independent velocity 
and  its transport is controlled by the dynamics of the 
cluster. In the main region $V_>$  (note that $V_>/(V_<+V_>) \sim 90 \%$)
the opposite limit $\tau_{ee}\ll\tau_>$ is realized and 
therefore the electron fluid is an independent entity and couples 
to the clusters by viscous friction. 
This situation is illustrated in Fig. 3.

The value of the 
viscous friction (or mutual friction) coefficient is related 
to $\tau_>$ as $\eta =(n_e/n_n)(\epsilon_{eF}/c^2\tau_>)$, 
where $n_e$ is the electron density. The values of the 
spin-up time (Table 2), which are identified with 
the dynamical coupling times of the superfluid shell at the crust-core
interface (Paper II),  are compatible with the requirement  
placed by the Christmas glitch observation of the Vela pulsar.

\section{Conclusions}

The main features of the present model and the predictions 
that can be drawn from these are as follows:

\begin{enumerate}
\item The glitch activity depends on the geometry of the 
crustal and core magnetic fields. If the generated field is 
anti-parallel to the spin axis (as predicted by the generation 
scenario) the inclination of the crustal field larger than 
90$^o$ would be necessary in order to produce a repulsive 
barrier to peripheral neutron vortices. If the crustal 
field geometries are restricted, then correlations 
between the glitch activity and the surface magnetic field values 
derived from (e.g.) the magnetic dipole radiation formula are expected.
If this is not the case, pulsars with  
similar characteristics (e.g. age) might show very different glitch activity.

\onecolumn
\begin{table*}
\begin{minipage}{155mm}
\begin{center}
\caption{Valuse of relaxation times and various microscopic parameters.}
\begin{tabular}{cccccccccc}
\hline
$\rho$ & $k_{Fe}$ & $\vert k \vert$ & $\delta_p$ & $\xi$ & $\tau_{ee}$&
$\tau_<$ & $\tau_>$ & $\eta$ & $\tau_d$ \\
$\times 10^{14}$ g cm$^{-3}$ & fm$^{-1}$ && fm & fm & $\times 10^{-15}$s  
& $\times 10^{-17}$ s&  s& g cm$^{-1}$
s$^{-1}$& min \\
\hline
1.67 &0.41 &0.14 &161.17 &6.57 &1.58 &3.85 &2.60e-09 &1.62e+15 &6.23e-01\\
1.84 &0.40 &0.18 &149.69 &7.01 &1.57 &3.48 &4.84e-10 &8.25e+15 &2.90+00\\
2.00 &0.46 &0.20 &140.05 &7.51 &1.88 &4.14 &1.43e-10 &4.50e+16 &1.45e+01\\
2.50 &0.52 &0.25 &118.08 &9.09 &2.32 &4.50 &1.75e-11 &6.11e+17 &1.58e+02\\
\hline
\end{tabular}
\end{center}
\end{minipage}
\end{table*}

\begin{figure}
\begin{center}
\mbox{\psfig{figure=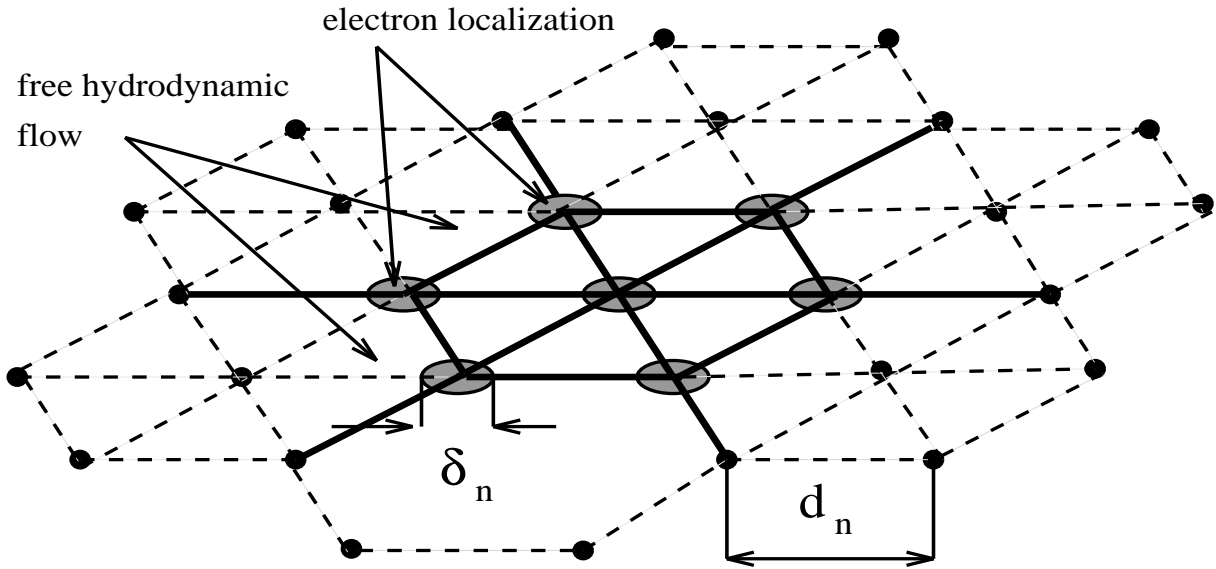,height=6.3in,width=3.in,angle=-90}}
\end{center}
\caption[] 
{\footnotesize{This figure shows the triangular neutron vortex lattice 
with the lattice constant $d_n$. The shaded  regions display the 
clusters of effective size $\delta_n$. The electrons form a 
hydrodynamic fluid in the intervening space between the cluster and 
are localized within the vortex clusters, providing a charge 
neutralizing background.}}
\label{fig3}
\end{figure}
\twocolumn
\noindent
This type of observation could support our model, because this
type of behaviour cannot be understood within known models of glitch
generation.
\item 
For millisecond pulsars with low magnetic fields ($\sim 10^{9}$ G) the
vortex-interface interaction force would be 6 orders of magnitude 
lower than for conventional pulsars ($H_0 \sim 10^{12}$ G), 
implying interjump  periods comparable to the evolutionary time-scales and, 
therefore, a marginal glitch activity.

\item The occurrence of the glitch would require small 
changes in the geometry of the compound field of the pulsar and 
possibly changes in the inclination angle of 
the dipole field. The location of the 
barrier at a fixed interface (with the same physical parameters for 
the whole system) and the homogeneity of the environment of the 
superfluid phase suggests consistency in the conditions at 
the onset of instability driving 
the glitch, which appears to be one of the advantages of the present model.

\item The thermal effects do not appear as an input;  
the relase of the self-energy of the vortex lines at the crust-core 
boundary might, however, provide a considerable heat flux (and the stationary 
situation is discussed in Sedrakian \& Sedrakian 1993).  
Thermal pulse observation associated 
with a glitch  
(Eichler \& Cheng 1989, Van Riper, Epstein \& Miller 1991, 
Chong \& Cheng 1994,  Hirano et al 1997) can be caused 
by the vortex annihilation at 
the crust core boundary. This type of heating may serve as an 
input for the thermally activated glitches (Link \& Epstein 1996).

\item The model does not invoke any type of quake activity, 
      but may trigger 
      magnetic field configuration changes, which may 
      induce plate tectonic activity 
      (Ruderman 1991). The opposite may be true, i.e. the 
      plate tectonic activity 
      (Ruderman 1991) could allow  the proton vortices to overcome 
      the barrier and trigger a runaway instability  at the 
      the crust core interface.
\end{enumerate}

To summarize, a potential barrier, resulting from magnetic interaction 
between the proton vortices and the crustal magnetic field 
at the crust-core interface, is suggested as the possible 
mechanism for the generation of pulsar  
glitches. It is shown that the force derived from that potential 
can sustain a Magnus force in an interjump period within a superfluid 
region with sufficient moment of inertia and appropriate dynamical 
coupling times in order to generate a large glitch, like those observed 
in the Vela-type pulsars. 

\section*{Acknowledgments}
This work was supported by NSF Grant 9528394 to Cornell University.
AS gratefully acknowledges a research grant from the Max Kade Foundation, 
New York.

\appendix\section{Ground State Vortex Configurations}

In this appendix we recapitulate several results 
regarding magnetic structure of the neutron vortex lattice. 
These has been previously obtained by Vardanian \& Sedrakian (1981) 
and  Sedrakian et al (1983) in a hydrodynamic approach and 
by Alpar, Langer \& Sauls (1984) using an  effective Ginzburg-Landau theory.
Here we employ a variational minimization method, which allows us to 
illuminate several aspects of the problem which have not been exposed previously.

Consider a rotating  Fermi-liquid mixture of superfluid 
neutrons, superconducting protons and normal 
electrons. At temperatures of interest 
the number of quasiparticle excitations is negligible. 
The electron system executes a rigid body rotation with angular 
velocity $\bmath\Omega$.
The kinetic energy of the neutron superfluid, 
which is by far the dominant part of 
the energy of the system,  is minimized by a lattice
of neutron vortices. This allows for a coarse-grained rigid body 
rotation of the superfluid with the angular velocity  ${\bmath\Omega}$.
If the interaction between neutrons and protons is switched on, the 
entrainment effect does not changes this result to any 
considerable extent: the correction to the mass current of the neutrons is 
of the order of the ratio of proton to neutron density, while 
the resulting magnetic energy density of the system is by orders of 
magnitude  lower than the kinetic energy density. Below we shall
assume that the neutron superflow pattern is 
determined by the minimization of 
the kinetic energy of this system and  consider
the behaviour of the proton superconductor.

The gradient invariant superfluid velocities are defined 
as \footnote{Note that in the rotating frame all velocities 
acquire a constant $\bmath\Omega \times \bmath r$ term, 
which however does not affect the result of variational calculation 
and therefore will be omitted everywhere below.} 
\be\label{1a} 
\vv_1 &=& \frac{\hbar}{2m_1} \bmath \nabla\chi_1 - \frac{e}{m_1c}\bmath A ,\\
\label{1b}
\vv_2 &=& \frac{\hbar}{2m_2} \bmath \nabla\chi_2 ,
\ee
where $m$ denotes the mass, $\chi$  the phase of the
superfluid order parameter and $\bmath A$  the vector potential;
the isospin indices  $1$ and $2$ refer to protons and neutrons 
respectively. Taking the curl of equations (\ref{1a}) and (\ref{1b}) and 
accounting for the quantization of the phase of superfluid order 
parameter, one finds 
\be\label{2a} 
\curl \vv_1 &=& \bmath\nu_1 \frac{\pi\hbar}{m_1} n_1 - \frac{e}{m_1c}\B, \\
\label{2b}
\curl \vv_2 &=& \bmath\nu_2 \frac{\pi\hbar}{m_2} n_2
\ee
where $n$ stands for the number density of vortices, $\bmath\nu=(\curl
\vv)\, /\vert \curl \vv \vert$ and $\B=\curl\bmath A$. The Maxwell equation 
for the magnetic field is   
\be\label{3} 
\curl \B = \frac{4\pi e}{m_1c}\left(\rho_{11}\vv_1 +\rho_{12}\vv_2\right),
\ee
where $\rho_{11}$ and $\rho_{12}$ are unentrained and entrained parts of 
the proton condensate density. We define auxiliary functions $\B_1$ and $\B_2$
by the following relations:
\be\label{4a} 
\curl \B_1 &\equiv &\frac{4\pi e}{m_1c}\rho_{11}\vv_1 ,\\
\label{4b}
\curl \B_2 &\equiv &\frac{4\pi e}{m_2c}\rho_{12}\vv_2.
\ee
According to eq. (\ref{3}) $\B = \B_1 + \B_2$. Taking the curl of 
equations (\ref{4a}) and (\ref{4b}) and using equations (\ref{2a}) and 
(\ref{2b}) one finds 
\be\label{4c} 
\curl\curl \B_1 & =& \frac{4\pi e}{m_1c} 
\rho_{11}~\left( \bmath\nu_1\frac{\pi\hbar}{m_1} n_1-\frac{e}{m_1c}\B\right)
\nonumber \\
& = &\delta_p^{-2}\left( \bmath\nu_1\Phi_0  n_1  -\B,\right) \\
\label{4d}
\curl\curl \B_2 &=& \bmath\nu_2\frac{4\pi e}{m_2c}\rho_{12}~ 
\frac{\pi\hbar}{m_2} n_2=\bmath\nu_2\delta_p^{-2}\Phi_1 n_2,
\ee
where 
\be 
\delta_p^2 = \frac{m_1^2c^2}{4\pi e^2\rho_{11}},\quad 
\Phi_0 = \frac{\pi\hbar c}{e},\quad \Phi_1 =\frac{\rho_{12}}{\rho_{11}} 
\Phi_0.
\ee
The free-energy of the superfluid neutron-proton mixture reads 
\be\label{5} 
{\cal F} = \frac{1}{2} \int \left(\rho_{11}v_1^2 + 2\rho_{12}
\vv_1\cdot\vv_2 + \rho_{22}v_2^2 \right) dV + \int \frac{B^2}{8\pi}dV.
\ee
The free energy is a functional of the neutron and proton superfluid
velocities. Once the minimization with respect of the neutron superfluid
velocity is performed, the proton superflow pattern is determined by the 
minimum of this functional with respect to $\vv_1$ at
constant $\vv_2$. The respective variation of the functional (\ref{5}) is
\be\label{6} 
\delta{\cal F} &=&
\int \left(\rho_{11}\vv_1 + \rho_{12}\vv_2 \right)\cdot \delta\vv_1 dV
\nonumber \\
&+& \frac{1}{4\pi}\int\left(\B_1+\B_2\right)\cdot \delta\B_1 dV .
\ee
Alternatively, using equation (\ref{4a}) to eliminate $\vv_1$ in favor
of $\B_1$, one finds
\be\label{7} 
\delta{\cal F} &=& \frac{\delta_p^2}{4\pi}
\int\left(\curl\B_1 + \curl \B_2 \right)\cdot 
\curl\delta\B_1 \nonumber \\ 
&+&
\frac{1}{4\pi}\int\left(\B_1+\B_2\right)\cdot \delta\B_1 dV \nonumber \\
&=& \frac{1}{4\pi}\int\left[\B_1 + \B_2 
+\delta_p^2\curl\curl~\left(\B_1+\B_2\right)\right]\cdot \delta\B_1 dV .
\nonumber \\
\ee
Here, to obtain the third line, we used the relation 
$\curl \bmath a\cdot \bmath b - \bmath b\cdot \curl \bmath a = 
\div [\bmath b\times \bmath a]$ and the assumption that 
surface integral  over the boundary of the system vanishes (note that 
this is not true at the crust-core interface!) As $\delta\B_1$
is arbitrary, the minimum (more precisely extremum)
condition implies that the expression in the 
bracket must vanish identically. Using equations (\ref{4c}) and (\ref{4d})
this condition takes the simple form 
\be \label{8}
\Phi_1 n_2 +(\bmath\nu_1 \cdot \bmath\nu_2)~\Phi_0  n_1 = 0
\ee
or, apart from the situation when   $\bmath\nu_1 \cdot \bmath\nu_2=0$,
\be 
 n_1  =  \frac{\vert k\vert}{\bmath\nu_1 \cdot \bmath\nu_2} n_2.
\ee
(Note that the entrainment coefficient is negative). This result shows that 
the actual flux of the neutron vortex is $\Phi_1\pm q\Phi_0$, where $q$ is 
an integer different form zero, which is in 
contrast to previous general knowledge.

The derivation above has a definite subtlety, because we ignored the
fact that  for  the superconducting proton { subsystem} 
not only is the velocity field $\vv_2$ fixed, but according to (\ref{4b}) the 
field component $\B_2$ is fixed as well. This fact can be visualized
by taking the limit of perfect entrainment ($\vert k \vert \to 1$).  
In this idealized case 
the whole proton superconductor follows the neutron vortex circulations
and the corresponding magnetic field is unscreened 
(its logarithmic divergence 
is cut-off by the finite size of the system).

If the entrainment is imperfect, then part of 
the superconducting proton condensate is available either for Meissner
screening the field or squeezing it into flux tubes.
The presence of the background field $\B_2$ implies that one should seek
the  minimum of the Gibbs thermodynamical 
potential (${\cal F}-\B\cdot\B_2/4\pi$):
\be\label{10}  
{\cal G} &= &\frac{1}{2} \int\left(\rho_{11}v_1^2 + 2\rho_{12} 
\vv_1\cdot\vv_2+ \rho_{22}v_2^2\right) dV \nonumber \\
&+& \int \frac{B^2}{8\pi}dV
-\int \frac{\B\cdot\B_2}{4\pi}dV. 
\ee 
The variation of the Gibbs free-energy (which goes in a full analogy with 
the foregoing discussion for $\delta{\cal F}$)  leads to the condition
\be  
(\Phi_1 n_2-B_2)  +(\bmath\nu_1 \cdot \bmath\nu_2)~\Phi_0  n_1 = 0,
\ee
or
\be  
(\bmath\nu_1 \cdot \bmath\nu_2)~ n_1 
=  \frac{(B_2-n_2\Phi_1)}{\Phi_0} . 
\ee 
For a region far from the neutron vortex one finds
\be 
n_1 = \frac{B_2}{\Phi_0},
\ee 
where we used the fact that the configuration $(\bmath\nu_1 \cdot \bmath\nu_2)
= 1$ would require the smallest value of $n_1$. The value of $B_2$ is
determined by integration of eq.  (\ref{4b}) for a given  
neutron superflow $\vv_2$. 
In particular,  for a single neutron vortex 
this would be the familiar  superfluid pattern which 
falls off from the centre of the  neutron vortex as $\sim 1/r$.
The value of $B_2$ turns out to be $\sim 10^{14}$ G and  
the resulting vortex number density  $n_1\sim 10^{20}$. The net proton vortex
number per neutron vortex is $\sim 10^{13}$ (Sedrakian et al 1983; 
Paper I).

\end{document}